\documentclass[letterpaper, 10 pt, conference]{ieeeconf}
\IEEEoverridecommandlockouts
\usepackage{hyperref}
\hypersetup{
  hidelinks 
}
\usepackage{bm, amssymb}
\usepackage{amsmath}
\usepackage[T1]{fontenc}
\usepackage[latin1]{inputenc}
\usepackage{url}
\usepackage{cite}
\usepackage{amsfonts}
\usepackage{graphicx}
\usepackage{epsfig}
\usepackage{multirow}

\newtheorem{example1}{Example}[section]

\newtheorem{theorem}{Theorem}

\newtheorem{lemma*}{Lemma}

\begin{document}
\title{\textbf{Towards a MATLAB Toolbox to compute backstepping kernels\\ using the power series method}}

\author{{Xin Lin,  Rafael Vazquez and Miroslav Krstic}
\thanks{Xin Lin is with the Department of Research Center of Satellite Technology, Harbin Institute of Technology, Harbin, 150080 China, e-mail: linxin29b@hit.edu.cn.}
\thanks{R. Vazquez is with the Department of Aerospace Engineering, Universidad de Sevilla, Camino de los Descubrimiento s.n., 41092 Sevilla, Spain, e-mail: rvazquez1@us.es.}%
\thanks{M. Krstic is with the Department of Mechanical and Aerospace Engineering, University of California San Diego, La Jolla, CA 92093-0411, USA.}%
}

\maketitle
 \interdisplaylinepenalty=2500

\baselineskip=.95 \normalbaselineskip



\begin{abstract}
In this paper, we extend our previous work on the power series method for computing backstepping kernels. Our first contribution is the development of initial steps towards a MATLAB toolbox dedicated to backstepping kernel computation. This toolbox would exploit MATLAB's linear algebra and sparse matrix manipulation features for enhanced efficiency; our initial findings show considerable improvements in computational speed with respect to the use of symbolical software without loss of precision at high orders. Additionally, we tackle limitations observed in our earlier work, such as slow convergence (due to oscillatory behaviors) and non-converging series (due to loss of analiticity at some singular points). To overcome these challenges, we introduce a technique that mitigates this behaviour by computing the expansion at different points, denoted as localized power series. This approach effectively navigates around singularities, and can also accelerates convergence by using more local approximations. Basic examples are provided to demonstrate these enhancements. Although this research is still ongoing, the significant potential and simplicity of the method already establish the power series approach as a viable and versatile solution for solving backstepping kernel equations, benefiting both novel and experienced practitioners in the field. We anticipate that these developments will be particularly beneficial in training the recently introduced neural operators that approximate backstepping kernels and gains.
\end{abstract}

\section{Introduction}
Backstepping has emerged as a powerful, versatile and broadly applicable technique for the control of systems governed by partial differential equations (PDEs).  Initially introduced for the design of feedback laws and observers in one-dimensional reaction-diffusion PDEs~\cite{krstic}, it has been extended to higher dimensions~\cite{nball} and applied to a remarkably diverse array of applications, including flow control~\cite{vazquez,vazquez-coron}, thermal convection loops~\cite{convloop}, thermoacoustic instabilities~\cite{rijke},  hyperbolic 1-D systems~\cite{vazquez-nonlinear,florent,krstic3},  multi-agent deployment~\cite{jie}, and wave equations~\cite{krstic2}. Backstepping even allows for closed-form control laws in certain cases~\cite{Vazquez2014} and even the design of adaptive controllers~\cite{krstic4}.

The implementation of backstepping controllers and observers hinge on the solution of associated kernel PDEs. These are typically linear hyperbolic equations (first- or second-order) defined on a triangular (Goursat~\cite{holten}) domain and subject to non-standard boundary conditions. The numerical solution of these equations presents a unique challenge, especially for complex systems or those with discontinuous kernels, a common feature in coupled systems (both parabolic~\cite{coupled-parabolic} and hyperbolic~\cite{coupled-hyperbolic}); these discontinuities should be accurately captured with the numerical approximation if the implementation aims to be precise. While this topic is critical to backstepping applications, surprisingly, it has not received focused attention in the research literature.

Existing approaches for solving kernel equations include finite difference methods~\cite{krstic,simon, deutscher,aamo}, symbolic successive approximation series~\cite{convloop}, or numerical solutions of integral forms of the equations~\cite{jad,auriol2}.  While some numerical methods exist for Goursat problems~\cite{day}, they have not yet been explored in the context of backstepping kernels. These techniques can often lack flexibility, particularly when adapting to specific kernel equations or handling discontinuities.

The use of power series as an alternative method for solving backstepping kernel equations was introduced in several works.  It was first proposed in~\cite{ascencio}, where convex optimization ideas were used to approximate kernels while maintaining stability, but without detailed analysis of convergence. Later, the method was used without proof in~\cite{leo} for a problem involving coupled parabolic equations, where piecewise-analytic kernels required the use of several series to account for discontinuities. The first rigorous proof of uniqueness and convergence was given in~\cite{jing-paper}, in the context of a multi-dimensional case with singularities at the origin, thus demonstrating both convergence and the kernel's existence. Finally,  in our recent work~\cite{Power-Series}, we presented the power series-based method as a general approach, providing a solid theoretical foundation leveraging the classical successive approximation proof. We also introduced a framework for kernel computation based on Mathematica's symbolic algebra capabilities. The advantages of the method (ease of implementation, accuracy, and the ability to produce parametric kernels) were highlighted, along with the primary limitation: the requirement of analyticity for system coefficients (satisfied by most backstepping applications).  Slow convergence in some cases might necessitate high-order approximations, although modern computing power mitigates this issue at least for simple cases.

In this work, we significantly advance our previous power series method for backstepping kernel computation by developing a MATLAB-based implementation.  Leveraging MATLAB's sparse matrix manipulation and linear algebra features, we achieve substantial computational efficiency gains (of several orders of magnitude) while preserving the precision of our symbolic solver. This transition marks an initial step towards a user-friendly toolbox for the controls community, as reflected in the name given to this paper. The paper provides links to Live Scripts prepared by the authors, so that interested readers can already try and implement by themselves the examples provided in the paper, or even adapt the code to cases of interest to them.

Furthermore, we address limitations encountered in our earlier work, such as slow or non-convergent series due to singularities as shown in examples of~\cite{Power-Series}.  We introduce a technique  employing power series expansions calculated at strategically chosen points. This approach effectively helps in circumventing singularities and accelerates convergence through \emph{localized} approximations, and this is directly shown by revisiting a problematic example of~\cite{Power-Series}.

These advancements, combined with the inherent simplicity of the power series method, establish it as a powerful and versatile tool for solving backstepping kernel equations. The MATLAB implementation further enhances its accessibility and efficiency. This is particularly relevant within the emerging context of training neural operators for backstepping kernel approximation, where solution quality plays a paramount role. Current progress in this area demonstrate the enormous potential of neural operators to effectively learn backstepping gains and controllers while maintaining stability guarantees. This is reflected in the latest works, which include methods for bypassing traditional gain computations entirely \cite{bhan2023neural}, designing neural operator-based observers and controllers \cite{krstic2023neural}, extending the approach to more complex hyperbolic PDE systems \cite{wang2023backstepping}, developing adaptive neural operator-based control designs \cite{lamarque2024adaptive}, and even implementing gain scheduling techniques using neural operators \cite{lamarque2024gain}. The methodology presented in this paper can significantly accelerate the generation of large solution sets required for training the approximating DeepONets used in these works, thereby facilitating their adoption and further advancement in the field of neural operator-based control for backstepping systems.

The focus of this paper is twofold. Firstly, we present our current progress on migrating the computational framework from Mathematica to MATLAB. To this end, in Section~\ref{sect-MATLAB}, the paper first focuses on a framework to pose the power series computation of an example problem as an (sparse) linear algebra problem easily solvable with MATLAB, and in Section~\ref{sect-examples} we review all the basic examples from our initial paper. Secondly, we analyze in Section~\ref{sect-expansionpoint} an idea to overcome the main weakness of the method, namely slowly-convergent or even non-convergent series, by moving the point of expansion, getting farther away from singular points; an example shows how both ideas are blended together to provide a much improved solution both in terms of computational efficiency and by avoiding non-convergent behaviour. MATLAB code is provided separately by the authors in the form of Live Scripts.
We finish in Section~\ref{sect-conclusions} with some concluding remarks and future lines of work.

\section{Towards a MATLAB toolbox}~\label{sect-MATLAB}
Our previous paper~\cite{Power-Series} introduced the concept of a power series solution by an example. We use the same example to illustrate the basic framework developed for the MATLAB solver, even if at the cost of some redundancy, to make this paper self-contained. 

\subsection{A review of the basic results of~\cite{Power-Series}}

Consider the reaction-diffusion equation,
\begin{eqnarray} \label{eqn-urd}
    u_t&=&\epsilon u_{xx}+\lambda(x) u,
\end{eqnarray}
for $t>0$, with $\epsilon>0$ and $\lambda(x)$ an analytic function in the domain $x\in[0,L]$, and with boundary conditions, 
\begin{eqnarray}\label{eqn-urdbc1}
    u(t,L)&=&U(t),\quad
    u(t,0)=0,\label{eqn-urdbc2}
\end{eqnarray}
where $U$ is the actuation variable. For sufficiently large $\lambda(x)>0$, (\ref{eqn-urd})--(\ref{eqn-urdbc2}) becomes open-loop unstable.

Applying backstepping~\cite{krstic}, one chooses a tuning parameter $c\geq0$; then, the stabilizing control law is
\begin{eqnarray}\label{eqn-U1}
U&=&\int_{0}^L K(L,\xi) u(\xi)d\xi,
\end{eqnarray}
where the kernel function $K(x,\xi)$
verifies the so-called \emph{kernel equations}:	 
\begin{eqnarray}
    K_{xx}(x,\xi)- K_{\xi \xi}(x,\xi)&=&
	 \frac{\lambda(\xi)+c}{\epsilon} K(x,\xi) \label{eqn-K_PDE}
\end{eqnarray}
with the boundary conditions
\begin{eqnarray}
    K(x,x)&=& -\frac{1}{2\epsilon} \int_0^x \left(\lambda(\xi)+c\right) d\xi \label{eqn-K_BC1} \\
    K (x,0)&=&0 \label{eqn-K_BC2}
\end{eqnarray}
in the \emph{triangular} domain $\mathcal T=\{(x,\xi):0 \leq \xi \leq x \leq L\}$.

Based on the power series method~\cite{Power-Series} (Theorem 1), we know there exists a kernel function $K(x,\xi)$, analytic in both variables, which can therefore be expressed by using a  \emph{double} power series,
\begin{equation}
	K(x,\xi)=\sum_{i=0}^{\infty} \sum_{j=0}^i K_{ij}x^{i-j} \xi^{j}    \label{eqn-series}
\end{equation}
where  $K_{ij}$ denotes the series coefficients.

To be a bit more precise about the convergence of (\ref{eqn-series}), denote as $\mathcal D_L$ the complex-valued open disc centered at the origin and of radius $L$, i.e., $\mathcal D_L=\{x \in \mathbb C:\vert x \vert < L\}$. The new domain where (\ref{eqn-series}) is guaranteed to be convergent is a polydisc $\mathcal D_{L+\delta} \times \mathcal D_{L+\delta}$ of $\mathbb C^2$, for some $\delta>0$; we require such $\delta$ since it is essential to evaluate the kernel at $x=L$ for the control law (\ref{eqn-U1}). The only requirement of Theorem 1 in~\cite{Power-Series} is that there exists $\delta>0$ such that $\lambda$ is analytic on $\mathcal D_{L+\delta}$.

In practice, an $N$-order truncation operator $T^N(\cdot)$ is defined for the power series to retain the terms of $x^i \xi^j$ with $i+j \leq N$ and discard the other terms, where $N$ indicates the series truncation order. Applying the $N$-order truncation operator to (\ref{eqn-series}) gives an $N$-order truncated approximation of the kernel function, 
\begin{equation}
    K^N(x,\xi) := T^N(K(x,\xi)) = \sum_{i=0}^{N} \sum_{j=0}^i K_{ij}x^{i-j} \xi^{j}   \label{eqn-seriesApprox}
\end{equation}

To compute (\ref{eqn-seriesApprox}), one starts by expressing the analytic function $[{\lambda(\xi)+c}]/{\epsilon}$ as a power series,
\begin{equation}
	\frac{\lambda(\xi)+c}{\epsilon} = \sum_{i=0}^{\infty} \lambda_i \xi^{i} \label{eqn-lambda-series}
\end{equation}
and then an $N$-order truncated approximation of $[{\lambda(\xi)+c}]/{\epsilon}$ is obtained by applying the $N$-order truncation operator,
\begin{equation}
    \frac{\lambda^N(\xi)+c}{\epsilon} := T^N \left(\frac{\lambda(\xi)+c}{\epsilon}\right) = \sum_{i=0}^{N} \lambda_i \xi^{i} \label{eqn-lambda-seriesApprox}
\end{equation} \\
Substituting (\ref{eqn-seriesApprox}) and (\ref{eqn-lambda-seriesApprox}) into (\ref{eqn-K_PDE})--(\ref{eqn-K_BC2}), and equating all terms having the same powers of $x$ and $\xi$, one can obtain equations with respect to the coefficient $K_{ij}$~\cite{Power-Series},
\begin{equation} \label{eqn-K_rec_series2}
	\begin{aligned}
		B_{(i-2)j} & = (i-j)(i-j-1) K_{ij} \\ 
			& -(j+2)(j+1) K_{i(j+2)}, \quad  0 \leq j \leq i-2
	\end{aligned}
\end{equation}
\begin{eqnarray}
	\sum_{j=0}^i K_{ij}  &=& -\frac{1}{2} \frac{\lambda_{i-1}}{i}, \quad  i\geq 1  \label{eqn-K_BC1_series2} \\
	K_{i0}&=&0, \quad \quad  i \geq 0 \label{eqn-K_BC2_series2}
\end{eqnarray}
where
\begin{equation}
	B_{ij}=\sum_{q=0}^j K_{(i-q)(j-q)} \lambda_{q} \label{def-Bij}
\end{equation}

\subsection{A MATLAB double series vector-matrix framework: definitions and operators}
	
Up to here, the development mimics the one given in our previous paper~\cite{Power-Series}, where the task of finding the coefficients is accomplished by using the symbolic capabilities of Mathematica. The symbolic procedure is fast enough for small orders, but the computation slows down when the size of equations becomes relatively high (which happens when the series truncation order $N$ is large). To overcome this difficulty, we propose a \emph{vector-matrix formulation }of the problem, to exploit MATLAB's linear algebra and sparse matrix manipulation features for much enhanced efficiency. Our basic framework is explained below.
	
The $N$-order approximated polynomial $K^N(x,\xi)$ of the kernel function given in (\ref{eqn-seriesApprox}) is written in a vector form as follows
\begin{equation} \label{eqn-K_vector} 
    K^N(x,\xi)=\sum_{i=0}^{N} \sum_{j=0}^i K_{ij} x^{i-j} \xi^{j} := {\bm \kappa}^{N} {\bm z}^N 
\end{equation}
where
\begin{equation}
    \left\{
    \begin{aligned}
    {\bm \kappa}^N &= [K_{00},\;K_{10},\;K_{11},\;\ldots,K_{N0},\;\ldots,K_{NN}] \\
    {\bm z}^N &= [x^0\xi^0,\;x^1\xi^0,\;x^0\xi^1,\;\ldots,x^N\xi^0,\;\ldots,x^0\xi^N]^T 
    \end{aligned} \right.
\end{equation}
It can be seen that $K_{ij}$ and $x^{i-j} \xi^{j}$ are respectively the $m(i,j)$-th term of ${\bm \kappa}^N$ and ${\bm z}^N$, i.e., ${\bm \kappa}^N (m(i,j))=K_{ij}$ and ${\bm z}^N (m(i,j))=x^{i-j} \xi^{j}$, where $m(i,j) := l(i-1)+j+1$, and where $l(i) := (i+1)(i+2)/2$. Thus, these functions $m$ and $l$ translate indexes of the double series to indexes of the coefficient vector.
	
The next step is to appropriately define transformation matrices so that the different operators that one needs to apply to $K^N(x,\xi)$ (truncation, differentiation, trace operators at the boundaries, multiplication by $\lambda(\xi)$) are transformed into a matrix operation. The general mathematical formulation of such operators is
\begin{equation} \label{eqn-TransMat}
    f(K^N(x,\xi)) = {\bm \kappa}^{N}  {\bm R}^{N}  {\bm z}^N
\end{equation}
where $f(\cdot)$ denotes any operator applied to $K^N(x,\xi)$, ${\bm R}^{N} \in \mathbb{R}^{l(N) \times l(N)} $ denotes the transformation matrix corresponding to the operator $f(\cdot)$. \eqref{eqn-TransMat} can be also written in the scalar form as
\begin{equation}  \label{eqn-TransMat_scalar}
    \begin{aligned}
    f(K^N(x,\xi)) & = {\bm \kappa}^{N}  {\bm R}^{N}  {\bm z}^N \\
    & = \sum_{i=0}^{N} \sum_{j=0}^i  \left( \sum_{k=1}^{l(N)}  {\bm r}_{m(i,j)}(k) {\bm \kappa}(k)  \right)  x^{i-j} \xi^{j}
  \end{aligned}
\end{equation}
where ${\bm r}_k$ denotes the $k$-th column of ${\bm R}^{N}$, and where $1 \leq k \leq l(N)$.

Based on \eqref{eqn-K_vector}, the operation of taking a first-order partial derivative of $K^N(x,\xi)$ with respect to the first variable $x$ is expressed as a truncated power series in the following way:
\begin{equation} \label{eqn-K_x}
    \begin{aligned}
    K^N_x(x,\xi)  & = \sum_{i=0}^{N} \sum_{j=0}^i (i-j) K_{ij}x^{i-j-1} \xi^{j} \\
    & = \sum_{i=0}^{N-1} \sum_{j=0}^i (i-j+1) K_{(i+1)j}x^{i-j} \xi^{j} \\
    & := {\bm \kappa}^{N} {\bm D}_x^{N} {\bm z}^{N}
    \end{aligned}
\end{equation}
where the matrix ${\bm D}_x^{N} \in \mathbb{R}^{l(N) \times l(N)}$. Comparing \eqref{eqn-K_x} and \eqref{eqn-TransMat_scalar}, one can obtain that the $m(i,j)$-th column of ${\bm D}_x^{N}$, ${\bm d}_{x,m(i,j)}$, is
\begin{equation} \label{eqn-D_x}
    {\bm d}_{x,m(i,j)} (k) = 
    \left\{
    \begin{array} {ll}
    i-j+1, & {\rm if} \;\; k=m(i+1,j) \\
    0, & \mathrm{otherwise}.  
    \end{array}  \right.
\end{equation}
where $0 \leq j \leq i \leq N-1$ and $1 \leq k \leq l(N)$. Similarly, the operation of taking the first-order partial derivative of $K^N(x,\xi)$ with respect to the second variable $\xi$ is written as a truncated power series by the expression
\begin{equation} \label{eqn-K_xi}
    \begin{aligned}
    K^N_{\xi}(x,\xi) & = \sum_{i=0}^{N} \sum_{j=0}^i j K_{ij}x^{i-j} \xi^{j-1} \\
    & = \sum_{i=0}^{N-1} \sum_{j=0}^i (j+1) K_{(i+1)(j+1)}x^{i-j} \xi^{j} \\
    & :={\bm \kappa}^{N} {\bm D}_\xi^{N} {\bm z}^{N}
    \end{aligned}
\end{equation}
where the matrix ${\bm D}_\xi^{N} \in \mathbb{R}^{l(N) \times l(N)} $. Comparing \eqref{eqn-K_xi} and \eqref{eqn-TransMat_scalar}, one can obtain the $m(i,j)$-th column of ${\bm D}_\xi^{N}$, ${\bm d}_{\xi,m(i,j)}$, as
\begin{equation} \label{eqn-D_xi}
    {\bm d}_{\xi,m(i,j)} (k) = 
    \left\{
    \begin{array} {ll}
    j+1, & {\rm if} \;\; k=m(i+1,j+1) \\
    0, & \mathrm{otherwise}.  
    \end{array}  \right.
\end{equation}
where $0 \leq j \leq i \leq N-1$ and $1 \leq k \leq l(N)$. Moreover, based on (\ref{eqn-K_x}) and (\ref{eqn-K_xi}), the second-order partial derivatives of $K^N(x,\xi)$ can be obtained as
\begin{equation} \label{eqn-k_xx_xixi_0}
	\left\{
	\begin{aligned}
		K^N_{xx}(x,\xi) &= {\bm \kappa}^{N} {\bm D}_{x}^{N} {\bm D}_{x}^{N} {\bm z}^{N} \\
		K^N_{\xi \xi}(x,\xi) &= {\bm \kappa}^{N} {\bm D}_{\xi}^{N} {\bm D}_{\xi}^{N} {\bm z}^{N}
	\end{aligned} \right.
\end{equation}
Note that the maximum power of $K^N_{xx}(x,\xi)$ and $K^N_{\xi \xi}(x,\xi)$ are both $N-2$, thus one can write
\begin{equation} \label{eqn-k_xx_xixi}
    \left\{
    \begin{aligned}
    K^N_{xx}(x,\xi) &= {\bm \kappa}^{N} {\bm D}_{xx}^{N-2} {\bm z}^{N-2} \\
    K^N_{\xi \xi}(x,\xi) &= {\bm \kappa}^{N} {\bm D}_{\xi \xi}^{N-2} {\bm z}^{N-2}
    \end{aligned} \right.
\end{equation}
where ${\bm D}_{xx}^{N-2}$ and ${\bm D}_{\xi \xi}^{N-2} $ include the first $l(N-2)$ columns of $ {\bm D}_x^{N} {\bm D}_x^{N}$ and ${\bm D}_\xi^{N} {\bm D}_\xi^{N}$, respectively. 

The truncation operator $T^r$ is simply defined as reducing the order of the series to $r$ (and the number of coefficients to $l(r)$). Then, substituting \eqref{eqn-series} into the left-side of the kernel equation \eqref{eqn-K_PDE} and applying the $(N-2)$-order truncation operator, one can obtain the following expression based on \eqref{eqn-k_xx_xixi},
\begin{equation}
    T^{N-2} \left( K_{xx}(x,\xi)- K_{\xi \xi}(x,\xi) \right) = 
    {\bm \kappa}^{N} ({\bm D}_{xx}^{N-2} - {\bm D}_{\xi \xi}^{N-2}) {\bm z}^{N-2}
\end{equation}

In addition, substituting \eqref{eqn-series} and \eqref{eqn-lambda-series} into the right-side  of the kernel equation \eqref{eqn-K_PDE} and rearranging the sum, one gets
\begin{equation}
    \frac{\lambda(\xi)+c}{\epsilon} K(x,\xi) = \sum_{i=0}^{\infty} \sum_{j=0}^i \left( \sum_{q=0}^j \lambda_{q} K_{(i-q)(j-q)}  \right) x^{i-j} \xi^{j}
\end{equation}
and applying the $(N-2)$-order truncation operator to the above equation yields
\begin{equation} \label{eqn-kernel_RHS}
    \begin{aligned}
    & T^{N-2} \left( \frac{\lambda(\xi)+c}{\epsilon} K(x,\xi) \right) \\ 
    & = \sum_{i=0}^{N-2} \sum_{j=0}^i \left( \sum_{q=0}^j \lambda_{q} K_{(i-q)(j-q)}  \right) x^{i-j} \xi^{j} \\
    & := {\bm \kappa}^N {\bm H}_{\xi,\lambda}^{N-2} {\bm z}^{N-2}
    \end{aligned}
\end{equation}
where the matrix ${\bm H}_{\xi,\lambda}^{N-2} \in \mathbb{R}^{l(N) \times l(N-2)} $. Comparing \eqref{eqn-kernel_RHS} and \eqref{eqn-TransMat_scalar}, one can obtain the $m(i,j)$-th column of ${\bm H}_{\xi,\lambda}^{N-2}$, ${\bm h}_{\xi,m(i,j)}$, as
\begin{equation} \label{eqn-H_lambda}
    {\bm h}_{\xi, m(i,j)} (k) = 
    \left\{
    \begin{array} {ll}
    \lambda_{q}, & {\rm if} \;\; k=m(i-q, j-q) \\
    0, & \mathrm{otherwise}.  
    \end{array}  \right.
\end{equation}
where $0 \leq q \leq j \leq i \leq N-2$ and $1 \leq k \leq l(N)$.

Based on \eqref{eqn-k_xx_xixi} and \eqref{eqn-kernel_RHS}, substituting \eqref{eqn-series} and \eqref{eqn-lambda-series} into the both sides of the kernel equation \eqref{eqn-K_PDE} and applying the $(N-2)$-order truncation operator, one can obtain
\begin{equation} \label{eqn-K_PDE_vec_z}
    {\bm \kappa}^N {\bm M}_{K} {\bm z}^{N-2} =  0
\end{equation}
where ${\bm M}_{K} = {\bm D}_{xx}^{N-2} - {\bm D}_{\xi \xi}^{N-2} - {\bm H}_{\xi,\lambda}^{N-2}$. Note that ${\bm \kappa}^N {\bm M}_K$ are the coefficients of a power series which equals zero, and therefore must be zero themselves. Thus one can write
\begin{equation} \label{eqn-K_PDE_vec}
    {\bm \kappa}^N {\bm M}_{K} = {\bm b}_{K} = {\bm 0}_{1 \times l(N-2)}
\end{equation}

For some operators $g(\cdot)$ that takes the second variable $\xi$ in $K^N(x, \xi)$ as a function of the first variable $x$ (e.g, the trace operator at the boundary in \eqref{eqn-K_BC1} which takes $\xi=x$), \eqref{eqn-TransMat_scalar} can be rewritten as
\begin{equation}  \label{eqn-TransMat_scalar2}
    g(K^N(x, \xi))  = {\bm \kappa}^{N}  {\bm S}^{N}  {\bm x}^N 
    = \sum_{i=0}^{N}  \left( \sum_{k=1}^{l(N)}  {\bm s}_{i+1}(k) {\bm \kappa}(k)  \right)  x^{i}
\end{equation}
where ${\bm x}^N=[x^0,\ldots,x^N]^T$, the matrix ${\bm S}^{N} \in \mathbb{R}^{l(N) \times (N+1)} $, and ${\bm s}_{i+1}$ indicates the $(i+1)$-th column of ${\bm S}^{N}$, where $0 \leq i \leq N$. Taking $\xi=\alpha x,\;\alpha \in \mathbb{R}$, and $K^N(x,\xi)$ given in \eqref{eqn-seriesApprox} can be expressed as
\begin{equation} \label{eqn-BC_LHS}
    K^N(x, \alpha x) = \sum_{i=0}^{N} \left( \sum_{j=0}^i \alpha^j K_{ij} \right) x^{i} := {\bm \kappa}^{N}  {\bm P}_\alpha^{N}  {\bm x}^N
\end{equation}
where the matrix ${\bm P}_\alpha^{N} \in \mathbb{R}^{ l(N) \times (N+1) } $. Comparing \eqref{eqn-BC_LHS} and \eqref{eqn-TransMat_scalar2}, one can obtain the $(i+1)$-th column of ${\bm P}_\alpha^{N}$, ${\bm p}_{i+1}$, as
\begin{equation} \label{eqn-BC_P}
    {\bm p}_{i+1} (k) = 
    \left\{
    \begin{array} {ll}
    \alpha^{k-m(i,0)}, & {\rm if} \;\; m(i,0) \leq k \leq m(i, i) \\
    0, & \mathrm{otherwise}.  
    \end{array}  \right.
\end{equation}
where $0 \leq i \leq N$ and $1 \leq k \leq l(N)$.

In addition, substituting (\ref{eqn-lambda-series}) into the right-side of (\ref{eqn-K_BC1}) and applying the $N$-order truncation operator, one can obtain
\begin{equation} \label{eqn-BC1_RHS}
    \begin{aligned}
    T^N \left( -\frac{1}{2\epsilon} \int_0^x \left(\lambda(\xi)+c\right) d\xi \right) 
    & = \sum_{i=0}^{N-1} \frac{-\lambda_i}{2(i+1)} x^{(i+1)} \\
    & := {\bm \lambda}_{\rm int}^N {\bm x}^N
    \end{aligned}
\end{equation}
where 
\begin{equation}
    {\bm \lambda}_{\rm int} = \left[0, \frac{-\lambda_0}{2},\dots,\frac{-\lambda_{N-1}}{2N}\right]
\end{equation}

Based on \eqref{eqn-BC_LHS} and \eqref{eqn-BC1_RHS}, substituting \eqref{eqn-series} and \eqref{eqn-lambda-series} into the boundary conditions \eqref{eqn-K_BC1} and \eqref{eqn-K_BC2} and applying the $N$-order truncation operator, one can obtain
\begin{eqnarray}
    \left( {\bm \kappa }^N {\bm P}_{1}^N - {\bm \lambda}_{\rm int}^N \right) {\bm x}^N  &=& 0 \\
    \left( {\bm \kappa }^N {\bm P}_{0}^N \right) {\bm x}^N   &=& 0
\end{eqnarray}
Similarly, note that ${\bm \kappa }^N {\bm P}_{1}^N - {\bm \lambda}_{\rm int}^N$ and ${\bm \kappa }^N {\bm P}_{0}^N$ are the coefficients of a power series which equals zero, and therefore must be zero themselves. Thus one can write
\begin{eqnarray}
    {\bm \kappa }^N {\bm P}_{1}^N   &=& {\bm \lambda}_{\rm int}^N  \label{eqn-K_BC1_0} \\
    {\bm \kappa }^N {\bm P}_{0}^N   &=& {\bm 0}_{1 \times (N+1)} \label{eqn-K_BC2_0}
\end{eqnarray}
\eqref{eqn-K_BC1_0} and \eqref{eqn-K_BC2_0} both represent $N+1$ equations about ${\bm \kappa }^N$. However, note that the first equations in \eqref{eqn-K_BC1_0} and \eqref{eqn-K_BC2_0} are the same, both being $K_{00} = 0$, thereby the first equation in \eqref{eqn-K_BC2_0} can be discarded. Therefore, \eqref{eqn-K_BC1_0} and \eqref{eqn-K_BC2_0} can be written as 
\begin{eqnarray}
    {\bm \kappa}^N {\bm M}_{BC,1}  &=& {\bm b}_{BC,1} = {\bm \lambda}_{\rm int}^N \label{eqn-BC1_vec} \\
    {\bm \kappa}^N {\bm M}_{BC,2}  &=& {\bm b}_{BC,2} = {\bm 0}_{1 \times N} \label{eqn-BC2_vec}
\end{eqnarray}
where ${\bm M}_{BC,1}={\bm P}_{1}^N$ and ${\bm M}_{BC,2}$ includes the second to the last column of ${\bm P}_{0}^N$.
	
Finally, based on \eqref{eqn-K_PDE_vec}, \eqref{eqn-BC1_vec}, and \eqref{eqn-BC2_vec}, the equations with respect to the series coefficient vector ${\bm \kappa }^N$ are obtained as
\begin{equation} \label{eqn-K_all}
    {\bm \kappa}^N {\bar {\bm M}} = {\bar {\bm  b}}
\end{equation}
where 
\begin{equation}
    \begin{aligned}
	{\bar {\bm M}} &= [{\bm M_{K}},{\bm M}_{BC,1},{\bm M}_{BC,2}]\\
	{\bar {\bm b}} &= [{\bm b_{K}},{\bm b}_{BC,1},{\bm b}_{BC,2}]
    \end{aligned}
\end{equation}
\eqref{eqn-K_all} is a system of linear equations with respect to ${\bm \kappa}^N$, and it is easy to verify that the numbers of unknown variables and equations are the same, both being $l(N)$. The linear equations of ${\bm \kappa}^N$ can be efficiently solved with MATLAB.

In addition, when the series truncation order $N$ becomes large, it can be seen from \eqref{eqn-D_x}, \eqref{eqn-D_xi}, and \eqref{eqn-H_lambda} that most elements of the transformation matrices ${\bm D}_x^{N}$, ${\bm D}_{\xi}^{N}$, and ${\bm H}_{\xi,\lambda}^{N-2}$ will be 0, which implies that the system matrix ${\bar {\bm M}}$ in \eqref{eqn-K_all} may be sparse. In this case, the sparse matrix manipulation features in MATLAB can be adopted to enhance the solving efficiency by reducing the memory occupied by these matrices.

\section{Basic examples revisited}\label{sect-examples}
Our previous paper~\cite{Power-Series} demonstrated the performance of the proposed power series method by some basic examples, which are solved by using the Mathematica solver~\cite{mathematica}. In this section, we revisit these examples with the developed MATLAB solver, except for a symbolic example. To avoid redundancy, we only repeat the kernel equations but not the plant equations and plot just the kernel gains and use different values of the coefficients except for Example 1 which is used for computational speed comparison. The examples can be downloaded as Live Scripts\footnote{The MATLAB Live Scripts can be downloaded from the website \url{http://aero.us.es/rvazquez/PowerSeriesLiveScripts.zip}.}, which are fully commented and can be reused by interested readers for their own kernel computations. Some examples in this section and the next provide computation times for comparison; all the tests are performed on an Intel$^\circledR$ Core$^\text{TM}$ i5-6300HQ 2.30GHz with Windows 10 and run on MATLAB 2018b.

\subsection{Example 1: Parabolic equation with space-varying reaction}
\begin{figure}[!t]
	\begin{centering}
		\includegraphics[width=6.5cm]{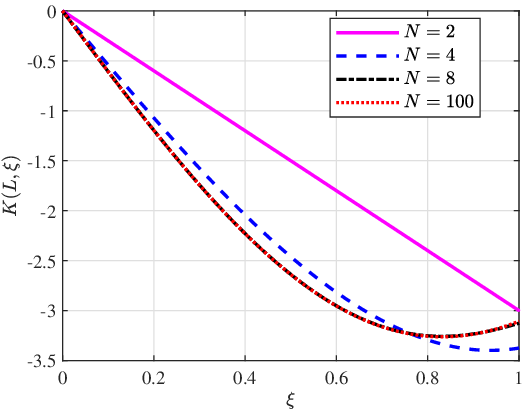}
		\caption{Convergent case with $\lambda(x)=3+x^2\sin(3x)$ for Example 1.}
		\label{fig-ex1a}
	\end{centering}
\end{figure}
	
{\renewcommand\baselinestretch{1.5}\selectfont
		\begin{table}[!t]
		\caption{Mathematica vs. MATLAB computation times (Example 1).} \vspace*{-0.5em}
		\scriptsize
			\label{Table_1a: computation times}
			\centering
			\begin{tabular}{lcccc}
				\hline\hline
				\multirow{3}{*}{$N$}&  \multicolumn{3}{c}{Computation time (s)}& 	 \multirow{3}{*}{Sparsity}\\ \cline{2-4}
				&    \multirow{2}{*}{Mathematica}&  \multicolumn{2}{c}{MATLAB}&   \\   \cline{3-4} 
				&   &    Non-sparse&    Sparse&  \\ \hline
				2&    2.03e-03&  6.20e-06& 5.64e-06& 0.694 \\
				4&    6.94e-03&  8.76e-06& 1.97e-05& 0.836 \\
				6&    1.66e-02&  1.98e-05& 3.11e-05& 0.895 \\
				8&    3.78e-02&  3.80e-05& 4.62e-05& 0.926 \\
				25&   2.43&  3.31e-03& 1.36e-03& 0.982 \\
				50&  176.49&  2.89e-02& 9.81e-03& 0.992 \\
				100&  ---&  8.72e-01& 6.58e-02& 0.996 \\
				\hline\hline
		\end{tabular}
\end{table}}
	
Consider $\lambda(x)=3+x^2\sin(3x)$, $L=\epsilon=1$, $c=3$. This verifies Theorem 1 in~\cite{Power-Series} and Fig.~\ref{fig-ex1a} shows the resulting kernel gain for several orders of approximation by using the proposed MATLAB solver. It can be seen from Fig.~\ref{fig-ex1a} that the obtained results by the proposed MATLAB solver are the same as those by the Mathematica solver in~\cite{Power-Series}. The computation times of the Mathematica solver and the MATLAB solver (with and without using sparse matrix definitions) for several series approximation orders are listed in Table~\ref{Table_1a: computation times}, where the results of the Mathematica solver for the case with $N=100$ is not provided since it takes too long. The results of Table~\ref{Table_1a: computation times} clearly show a very significant computational efficiency advantage of the MATLAB solver relative to the Mathematica solver, particularly for large orders. The Mathematica solver shows sufficient computational speed only when $N$ is small (e.g., $N \leq 8$), whereas the developed MATLAB solver is still fast enough even with a large $N$ (e.g., $N=100$). In addition, the sparsity of the problem for different series approximation orders are also listed in Table~\ref{Table_1a: computation times}, where the sparsity is defined as the ratio of the number of zero coefficients of the system matrix ${\bar {\bm M}}$ in \eqref{eqn-K_all} to the number of all elements of ${\bar {\bm M}}$. The results of Table~\ref{Table_1a: computation times} show that when the sparsity of the problem is high (e.g., the cases with $N=50$ and $N=100$), the use of the sparse matrix manipulation features in MATLAB can enhance the computational efficiency (as well as memory savings, not shown).
	
\subsection{Example 2: Parabolic equation with space-varying diffusion}
\begin{figure}[!t]
	\begin{centering}
		\includegraphics[width=6.5cm]{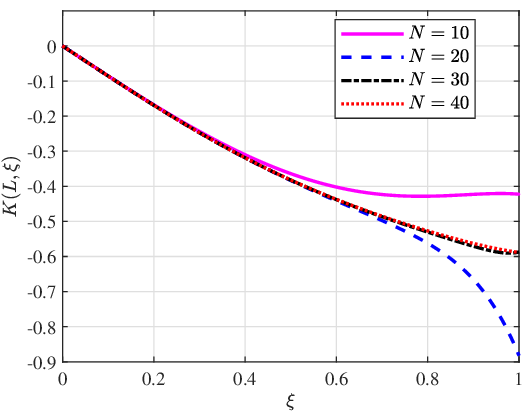}
		\caption{Convergent space-varying diffusion case,  $\lambda(x)=3+x\sin(6x)$, $\epsilon(x)=2+x^2$ (Example 2).}
		\label{fig-ex2a}
	\end{centering}
\end{figure}
	
This example was taken from~\cite{andrey} and was Example 3 in~\cite{Power-Series}. The kernel equations to be solved are:
\begin{eqnarray}
	\epsilon(x) K_{xx}(x,\xi)-\epsilon(\xi) K_{\xi \xi}(x,\xi)&=&
	(\lambda(\xi)+c) K(x,\xi)\quad\,\, \label{eqn-Keps1}
\end{eqnarray}
with $\epsilon(x)$ and $\lambda(x)$ analytic functions in $\mathcal D_L$, representing diffusion and reaction. The boundary conditions of the kernel equations are
\begin{eqnarray}
	2 \epsilon(x) \frac{d}{dx} K(x,x) &=& -\epsilon'(x) K(x,x)-\lambda(x)-c \\
	K (x,0)&=&0 \label{eqn-Keps3}
\end{eqnarray}
	
Consider $\lambda(x)=2+x^2\cos(6x^2)$, $\epsilon(x)=3+2x^3$, $L=1$, and $c=3$. This verifies Theorem 3 in~\cite{Power-Series} and Fig.~\ref{fig-ex2a} shows the resulting kernel gain for several orders of approximation by using the proposed MATLAB solver.

\subsection{Example 3: $2\times2$ 1-D linear hyperbolic system with space-varying coefficients}
\begin{figure*}[!t]
	\begin{centering}
		\includegraphics[width=6.25cm]{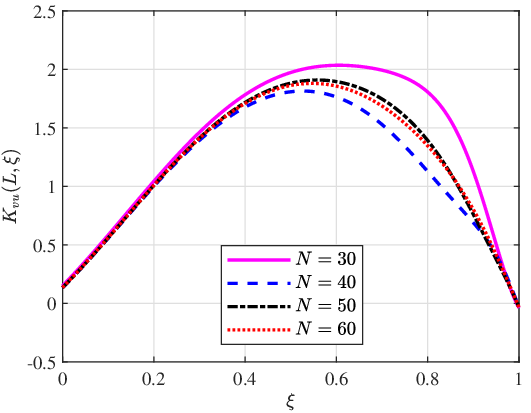}
		\includegraphics[width=6.25cm]{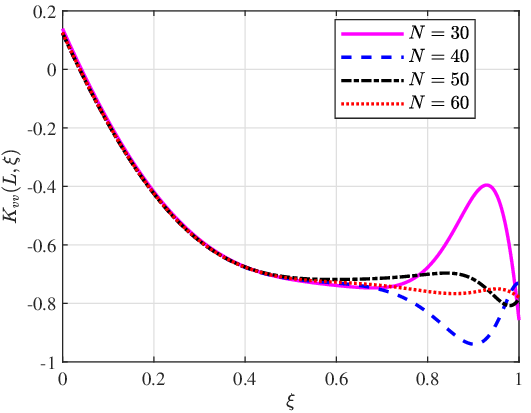}
		\caption{Hyperbolic 2x2 kernel gains $K_{vu}(L,\xi)$ (left) and $K_{vv}(L,\xi)$ (right) for different orders, showing convergence (Example 3).}
		\label{fig-ex3}
	\end{centering}
\end{figure*} 
	
	
This was Example 4 in~\cite{Power-Series}. In this case, backstepping requires finding the solution for two kernels, which we denote as $K^{vv}(x,\xi)$ and $K^{vu}(x,\xi)$ in the domain $\mathcal T$. The kernel equations are
\begin{eqnarray}\label{eqn-kvv}
	\mu(x)K_x^{vv}+\mu(\xi)K_\xi^{vv}&=&-\mu'(\xi)K^{vv}+c_2(\xi) K^{vu}
	\nonumber \\ && +\left[c_4(x)-c_4(\xi)\right] K^{vu},\quad\\
	\mu(x) K_x^{vu}-\epsilon(\xi) K_\xi^{vu}&=&\epsilon'(\xi)K^{vu}+
	c_3(\xi) K^{vv}
	\nonumber \\ &&+\left[c_4(x)-c_1(\xi)\right]K^{vv},\quad
\end{eqnarray}
with boundary conditions
\begin{eqnarray}
	K^{vv}(x,0)&=&\frac{q \epsilon(0)}{\mu(0)} K^{vu}(x,0),
	\label{eqn-bc3}\\
	(\epsilon(x)+\mu(x)) K^{vu}(x,x)&=&-c_3(x), \label{eqn-bc4}
\end{eqnarray}
Here, $\epsilon(x),\mu(x),c_i(x)$ are analytic in $\mathcal D_L$ and represent, respectively, the two transport speeds and the coupling coefficients.
Consider $\mu(x) = 1.4+x^3$, $\epsilon(x)=1.3+x^2$, $L = 1$, $
c_1(x) = 3\exp(3x)$,
$c_2(x)=\sin(3x)$,
$c_3(x)=1+2\cos(2x)$,
$c_4(x)=\frac{1}{3+1.5y^3}$, and 
$q=1$. This verifies Theorem 4 in~\cite{Power-Series} and Fig.~\ref{fig-ex3} shows the resulting kernel gain for several orders of approximation by using the proposed MATLAB solver.
	
\subsection{Example 4: Motion planning kernels for $(0+2)\times(0+2)$ 1-D linear hyperbolic system with space-varying coupling}
\begin{figure}[!b]
	\begin{centering}
		\includegraphics[width=6.5cm]{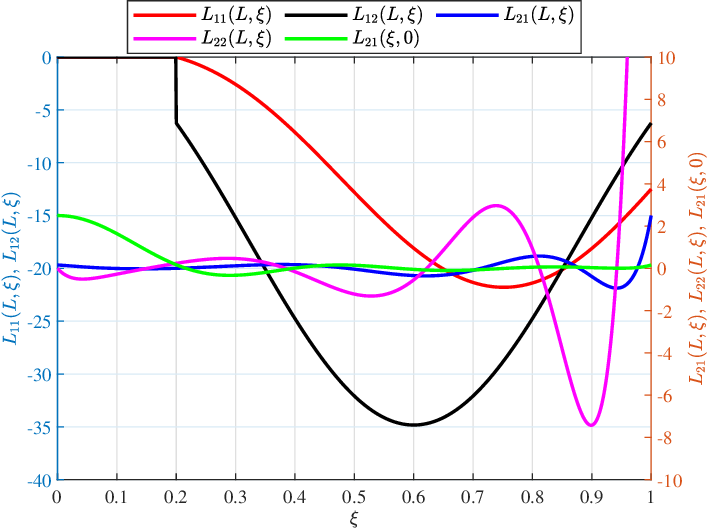}
		\caption{Solutions of gain kernels for Example 4. Note the discontinuous kernel $L_{12}(L,\xi)$ and the piecewise differentiable kernel $L_{11}(L,\xi)$.}
		\label{fig-ex4}
	\end{centering}
\end{figure} 
This was Example 5 in~\cite{Power-Series} and directly extracted from~\cite{coupled-hyperbolic}, having even a explicit solution for constant coefficients, which is the case under consideration. In this case one needs to find five gain kernels from $L_{11}$, $L_{12}$, $L_{21}$ and $L_{22}$, that verify
\begin{eqnarray} \label{eqn-ku1}
	\mu_1 \partial_x L_{11}(x,\xi)+\mu_1 \partial_\xi L_{11}(x,\xi)&=&
	\sigma_{21}(\xi)  L_{12}(x,\xi)\label{eqn-L11},\quad\\
	\mu_1 \partial_x L_{12}(x,\xi)+\mu_2 \partial_\xi L_{12}(x,\xi)&=&\sigma_{12} (\xi)L_{11}(x,\xi) ,\label{eqn-L12}\quad\\
	\mu_2 \partial_x L_{21}(x,\xi)+\mu_1 \partial_\xi L_{21}(x,\xi)&=&
	\sigma_{21}(\xi) L_{22}(x,\xi),\quad
	\\
	\mu_2 \partial_x L_{22}(x,\xi)+\mu_2 \partial_\xi L_{22}(x,\xi)&=&\sigma_{12} (\xi)L_{21}(x,\xi)
	,\quad
\end{eqnarray}
with boundary conditions
\begin{eqnarray}
	L_{11}(x,0)&=& L_{12}(x,0)=L_{22}(x,0)=0, \\
	L_{12}(x,x) &=&\frac{\sigma_{12}(x)}{\mu_2-\mu_1}, 
	L_{21}(x,x)=\frac{\sigma_{21}(x)}{\mu_1-\mu_2}. \label{eq:bcl21}
\end{eqnarray}
Here, $\mu_1>\mu_2>0$ are the (constant) transport speeds. As explained in~\cite{coupled-hyperbolic}, $L_{12}$, differently from the other kernels, possesses two boundary conditions, namely $L_{12}(x,0)=0$ and $L_{12}(x,x)=\frac{\sigma_{12}(x)}{\mu_2-\mu_1}$, which is solved by defining a piecewise smooth kernels; because of the coupling with $L_{11}$ this necessitates a similar piecewise definition for $L_{11}$.

Consider the particular case $\mu_1=1$, $\mu_2=0.2$, $\sigma_{12}=5$, and $\sigma_{21}=2$ considered in~\cite{coupled-hyperbolic}. The example presents numerical difficulties due to oscillations for these chosen coefficients and the Mathematica symbolic solver of~\cite{Power-Series} takes too long to obtain an accurate solution. To mitigate the problem, note that the kernels $(L_{11}, L_{12})$ and $(L_{21}, L_{22})$ can be computed individually, thereby the series truncation orders can be selected individually, which are denoted as $N_{L1}$ for $(L_{11}, L_{12})$ and $N_{L2}$ for $(L_{21}, L_{22})$. The resulting gain kernels are shown in Fig.~\ref{fig-ex4} for $N_{L1}=8$ and $N_{L1}=40$, which shows that a small order is enough for $(L_{11}, L_{12})$ and a relatively large order is needed for $(L_{21}, L_{22})$ (which are more oscillatory). The obtained results are consistent with those by the explicit solutions given in~\cite{coupled-hyperbolic} (see Fig. 1 in that paper), and the computation time is 9.34e-03 s.

\section{Changing the point of expansion to accelerate convergence and avoid singularities}~\label{sect-expansionpoint}
\begin{figure} [!t]
	\begin{centering}
		\includegraphics[width=6.5cm]{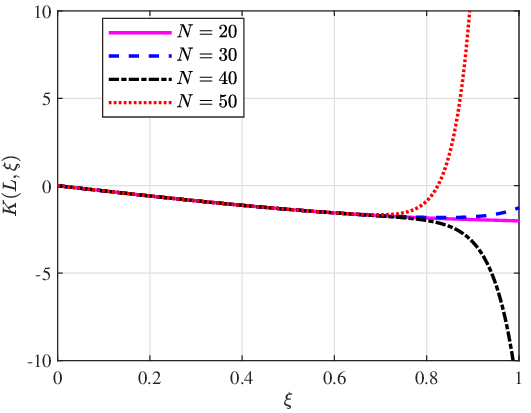}
		\caption{Divergent case with $\lambda(x)=\sqrt{0.5+x^2}$ (Example 1).}
		\label{fig-ex1b}
	\end{centering}
\end{figure}

For simplicity, the arguments of this section refer to the basic example (\ref{eqn-K_PDE})--(\ref{eqn-K_BC2}). According to Theorem 1 in~\cite{Power-Series}, a $\delta>0$ needs to exist such that $\lambda(x)$ is analytic on the disc $\mathcal D_{L+\delta}$ to ensure the convergence of the power series solution in \eqref{eqn-series}. When the function is not analytic within the disc $\mathcal D_{L+\delta}$, the power series solution diverges. For example, if one has $\lambda(x)=\sqrt{0.5+x^2}$ which is not analytic on the unit disc, the kernel gain does not converge as the order increases as it can clearly be seen in the Fig.~\ref{fig-ex1b} by inspecting the gain for different orders of the series. In this section, we will show that for some cases the divergent problem can be resolved by expanding (\emph{localizing}) the series at a different point. We call this new power series, expanded at a strategically-chosen point, a ``localized'' power series.
	
Now, the kernel  $K(x,\xi)$ is expressed by using a \emph{double} power series around the point $(x_0,\xi_0)$ rather than the origin, 
\begin{equation} \label{eqn-series_patch0}
	K(x,\xi)=\sum_{i=0}^{\infty} \sum_{j=0}^i K_{ij} (x-x_0)^{i-j} (\xi-\xi_0)^{j}    
\end{equation}
Denote $\tilde{x}=x-x_0$ and $\tilde{\xi}=\xi-\xi_0$. Then, \eqref{eqn-series_patch0} becomes
\begin{equation} \label{eqn-series_patch}
	\tilde{K}(\tilde{x}, \tilde{\xi})=\sum_{i=0}^{\infty} \sum_{j=0}^i K_{ij} \tilde{x}^{i-j} \tilde{\xi}^{j}
\end{equation}
Based on \eqref{eqn-series_patch}, the \emph{kernel equations} in \eqref{eqn-K_PDE} can be transformed into those in terms of $\tilde{x}$ and $\tilde{\xi}$,
\begin{equation} \label{eqn-K_PDE_patch}
    \tilde{K}_{\tilde{x} \tilde{x} }(\tilde{x}, \tilde{\xi})- \tilde{K}_{\tilde{\xi} \tilde{\xi}}(\tilde{x}, \tilde{\xi})= \frac{\lambda(\tilde{\xi}+\xi_0)+c}{\epsilon} \tilde{K}(\tilde{x}, \tilde{\xi})
\end{equation}
For the left-sides of boundary conditions in \eqref{eqn-K_BC1} and \eqref{eqn-K_BC2}, the values of the second variable $\xi$ are respectively $\xi = x$ for the first boundary condition and $\xi = 0$ for the second. Thus, in terms of $\tilde{x}$ and $\tilde{\xi}$, they result in $\tilde{\xi} = \tilde{x}+x_0-\xi_0$ for the first boundary condition and $\tilde{\xi} = -\xi_0$ for the second. For the integral in the right-side of boundary condition in \eqref{eqn-K_BC1}, it can be also expressed in terms of $\tilde{x}$ and $\tilde{\xi}$,
\begin{eqnarray} 
	&&\frac{-1}{2\epsilon} \int_0^x \left(\lambda(\xi)+c\right) d\xi \nonumber \\ &=& 
		\frac{-1}{2\epsilon} \int_0^{\tilde{x}+x_0} \left(\lambda(\tilde{\xi}+\xi_0)+c\right) d(\tilde{\xi}+\xi_0) \nonumber \\
		 &=&\frac{-1}{2\epsilon} \int_{-\xi_0}^{\tilde{x}+x_0-\xi_0} \left(\lambda(\tilde{\xi}+\xi_0)+c\right) d\tilde{\xi} \label{eqn-BC_RHS_patch}
\end{eqnarray}
Based on the change of variables and on \eqref{eqn-BC_RHS_patch}, the boundary conditions in \eqref{eqn-K_BC1} and \eqref{eqn-K_BC2} can be expressed in terms of $\tilde{x}$ and $\tilde{\xi}$, 
\begin{eqnarray}
	\tilde{K}(\tilde{x}, \tilde{x}_1) &=& \frac{-1}{2\epsilon} \int_{-\xi_0}^{\tilde{x}_1} \left(\lambda(\tilde{\xi}+\xi_0)+c\right) d\tilde{\xi} \label{eqn-K_BC1_patch} \\
		\tilde{K} (\tilde{x}, -\xi_0) &=& 0 \label{eqn-K_BC2_patch}
\end{eqnarray}
where $\tilde{x}_1=\tilde{x}+x_0-\xi_0$.
	
For the transformed kernel equations in \eqref{eqn-K_PDE_patch} and the boundary conditions in \eqref{eqn-K_BC1_patch} and \eqref{eqn-K_BC2_patch}, a Theorem similar to Theorem 1 in~\cite{Power-Series} can be stated for the given $x_0$ and $\xi_0$:
\begin{theorem}[Localized power series] \label{th-analytic_patch} 
    If there exists $\delta>0$ such that $\lambda(\xi)$ is analytic on the disc $\mathcal D_{L+\delta}(\xi_0)$ centered at $\xi_0$, and such disc covers the interval $[0,1]$, then there exists a $x_0$ and power series solution in the form of \eqref{eqn-series_patch0} which converges and defines an analytic function in the polydisc $\mathcal D_{L+\delta/2}(x_0) \times \mathcal D_{L+\delta/2}(\xi_0)$, that is the unique solution of the kernel equations (\ref{eqn-Keps1})--(\ref{eqn-Keps3}).
\end{theorem} 

The proof of the above Theorem is similar to Theorem 1 in~\cite{Power-Series}, since only the point of expansion varies and in~\cite{Power-Series} it was shown that the kernel is analytic in $\mathcal T$; by uniqueness of analytic functions any converging power series representing the kernel has to solve the kernel equations. In principle one can choose $x_0=\xi_0$ but other values are admissible. This theorem can be easily adapted to other examples, which is not done here due to lack of space. Instead, some numerical evidence is provided next for the basic example.

\begin{figure} [!t]
	\begin{centering}
		\includegraphics[width=6.5cm]{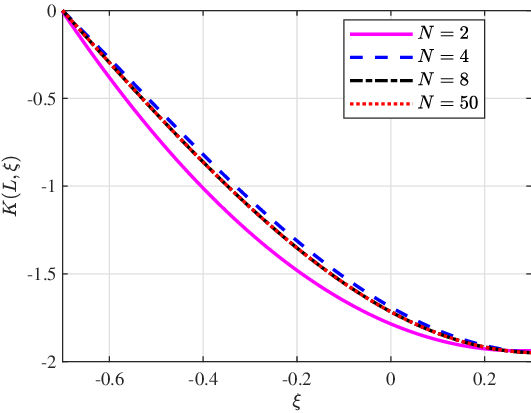}
		\caption{Convergent case with $\lambda(x)=\sqrt{0.5+x^2}$ by localized power series (Example 1).}
		\label{fig-ex1b_patch}
	\end{centering}
\end{figure}
	
{\renewcommand\baselinestretch{1.5}\selectfont
	\begin{table}[!t]
		\caption{Origin-based (divergent) vs. localized (convergent)\\ power series} \vspace*{-0.5em}
		\scriptsize
		\label{Table_1b: Sparsities and computation times}
		\centering
		\begin{tabular}{lccccc}
			\hline\hline
			\multirow{2}{*}{$N$}&  \multicolumn{2}{c}{Computation time (s)}& &  \multicolumn{2}{c}{Sparsity} \\ \cline{2-3} \cline{5-6}
			&    At origin&    Localized& & At origin&    Localized \\ \hline
			2&    5.70e-06&  6.47e-06& & 0.694& 0.556 \\
			4&    8.71e-06&  9.22e-06& & 0.831& 0.702 \\
			6&    2.42e-05&  2.00e-05& & 0.890& 0.783 \\
			8&    3.91e-05&  4.01e-05& & 0.921& 0.832 \\
			15&   2.41e-04&  4.66e-03& & 0.964& 0.908 \\
			25&   2.27e-03&  1.77e-03& & 0.981& 0.945 \\
                50&   3.53e-03&  1.52e-02& & 0.992& 0.973 \\
			\hline\hline
		\end{tabular}
\end{table}}

\subsection{Example 5: Parabolic equation with space-varying reaction not verifying Theorem 1 in~\cite{Power-Series}}
Choosing $x_0=0.5$ and $\xi_0=0.7$, it can be verified that the above Theorem is satisfied for $\lambda(x)=\sqrt{0.5+x^2}$, and the resulting kernel gain for several orders of approximation by using the proposed MATLAB solver is plotted in Fig.~\ref{fig-ex1b_patch}, which clearly shows that the localized power series solution is convergent. In addition, the sparsities and computation times of the zero-centered solution (divergent solution) and the localized solution (convergent solution) are listed in Table.~\ref{Table_1b: Sparsities and computation times}, which shows that convergence is achieved at the cost of losing some sparsity and computational speed; the loss is in any case reasonable.

The idea of a localized power series gives rise to an additional concept, that of \emph{patches} of power series, where instead of a single power series representation one employs several localized approximations. This would allow to better approximate oscillatory kernels while at the same time using smaller orders of approximation, for instance such an idea could reduce the order required to solve the Example 4 of Section~\ref{sect-examples}. This concept will be explored in future work.
 
\section{Concluding remarks}\label{sect-conclusions}
This paper presents significant advancements in the development of a MATLAB toolbox for computing backstepping kernels using the power series method, which combined with the simplicity of the method, establish it as a powerful and versatile tool for solving backstepping kernel equations. The MATLAB implementation enhances its accessibility and efficiency, making it particularly relevant for the emerging field of neural operator-based control for backstepping systems.

Future work will focus on the continued development of the MATLAB toolbox, incorporating additional features and optimizations to enhance its usability and performance. In particular the promising concept of patched power series (as explained at the end of Section~\ref{sect-expansionpoint}) will be explored. In addition, the application of the power series method to more complex systems, such as those with spatially-varying transport speed/diffusion coefficients and discontinuities defined by analytic differential equations, will also be explored. By addressing these challenges, the authors aim to expand the scope and applicability of the power series approach.

\section*{Acknowledgments}

Rafael Vazquez acknowledges support by grant TED2021-132099B-C33 funded by MICIU/AEI/10.13039 /501100011033 and by ``European Union NextGenerationEU/PRTR.'' The work is partially supported by the China Scholarship Council.

\bibliographystyle{IEEEtranS}  
\bibliography{references}

\begin{thebibliography}{10}
\providecommand{\url}[1]{#1}
\csname url@samestyle\endcsname
\providecommand{\newblock}{\relax}
\providecommand{\bibinfo}[2]{#2}
\providecommand{\BIBentrySTDinterwordspacing}{\spaceskip=0pt\relax}
\providecommand{\BIBentryALTinterwordstretchfactor}{4}
\providecommand{\BIBentryALTinterwordspacing}{\spaceskip=\fontdimen2\font plus
\BIBentryALTinterwordstretchfactor\fontdimen3\font minus
  \fontdimen4\font\relax}
\providecommand{\BIBforeignlanguage}[2]{{%
\expandafter\ifx\csname l@#1\endcsname\relax
\typeout{** WARNING: IEEEtranS.bst: No hyphenation pattern has been}%
\typeout{** loaded for the language `#1'. Using the pattern for}%
\typeout{** the default language instead.}%
\else
\language=\csname l@#1\endcsname
\fi
#2}}
\providecommand{\BIBdecl}{\relax}
\BIBdecl

\bibitem{rijke}
G.~Andrade, R.~Vazquez, and D.~Pagano, ``Backstepping stabilization of a
  linearized {ODE-PDE} {R}ijke tube model,'' \emph{Automatica}, vol.~96, pp.
  98--109, 2018.

\bibitem{aamo}
H.~Anfinsen and O.~Aamo, \emph{Adaptive control of hyperbolic {PDE}s}.\hskip
  1em plus 0.5em minus 0.4em\relax New York: Springer, 2019.

\bibitem{ascencio}
P.~Ascencio, A.~Astolfi, and T.~Parisini, ``Backstepping {PDE} design: A convex
  optimization approach,'' \emph{IEEE Transactions on Automatic Control},
  vol.~63, pp. 1943--1958, 2018.

\bibitem{auriol2}
J.~Auriol and D.~Bresch-Pietri, ``Robust state-feedback stabilization of an
  underactuated network of interconnected $n+m$ hyperbolic {PDE} systems,''
  \emph{Automatica}, vol. 136, p. 110040, 2022.

\bibitem{bhan2023neural}
L.~Bhan, Y.~Shi, and M.~Krstic, ``Neural operators for bypassing gain and
  control computations in {PDE} backstepping,'' \emph{arXiv preprint
  arXiv:2302.14265}, 2023.

\bibitem{leo}
L.~Camacho-Solorio, R.~Vazquez, and M.~Krstic, ``Boundary observers for coupled
  diffusion-reaction systems with prescribed convergence rate,'' \emph{Systems
  and Control Letters}, vol. 134, p. 104586, 2020.

\bibitem{vazquez-nonlinear}
J.-M. Coron, R.~Vazquez, M.~Krstic, and G.~Bastin, ``Local exponential ${H^2}$
  stabilization of a ${2\times 2}$ quasilinear hyperbolic system using
  backstepping,'' \emph{SIAM J. Control Optim.}, vol.~51, pp. 2005--2035, 2013.

\bibitem{day}
J.~Day, ``A {R}unge-{K}utta method for the numerical solution of the {G}oursat
  problem in hyperbolic partial differential equations,'' \emph{The Computer
  Journal}, vol. 9(1), pp. 81--83, 1966.

\bibitem{deutscher}
J.~Deutscher and J.~Gabriel, ``A backstepping approach to output regulation for
  coupled linear wave-{ODE} systems,'' \emph{Automatica}, vol. 123, p. 109338,
  2021.

\bibitem{florent}
F.~Di~Meglio, R.~Vazquez, and M.~Krstic, ``Stabilization of a system of n+1
  coupled first-order hyperbolic linear {PDE}s with a single boundary input,''
  \emph{IEEE Trans. Aut. Contr.}, vol.~58, pp. 3097--3111, 2013.

\bibitem{holten}
R.~Holten, ``Generalized {G}oursat problem,'' \emph{Pacific Journal of
  Mathematics}, vol.~12, pp. 207--224, 1962.

\bibitem{coupled-hyperbolic}
L.~Hu, F.~Di~Meglio, R.~Vazquez, and M.~Krstic, ``Control of homodirectional
  and general heterodirectional linear coupled hyperbolic {PDE}s,'' \emph{IEEE
  Transactions on Automatic Control}, vol.~61, no.~10, pp. 3301--3314, 2016.

\bibitem{jad}
L.~Jadachowski, T.~Meurer, and A.~Kugi, ``An efficient implementation of
  backstepping observers for time-varying parabolic {PDE}s,'' \emph{IFAC
  Proceedings Volumes}, vol.~45, pp. 798--803, 2012.

\bibitem{simon}
S.~Kerschbaum and J.~Deutscher, ``Backstepping control of coupled linear
  parabolic {PDE}s with space and time dependent coefficients,'' \emph{IEEE
  Trans. Aut. Contr.}, vol. 65(7), pp. 3060--3067, 2019.

\bibitem{krstic3}
M.~Krstic and A.~Smyshlyaev, ``Backstepping boundary control for first order
  hyperbolic {PDE}s and application to systems with actuator and sensor
  delays,'' \emph{Syst. Contr. Lett.}, vol.~57, pp. 750--758, 2008.

\bibitem{krstic}
------, \emph{Boundary Control of {PDE}s}.\hskip 1em plus 0.5em minus
  0.4em\relax SIAM, 2008.

\bibitem{krstic2023neural}
M.~Krstic, L.~Bhan, and Y.~Shi, ``Neural operators of backstepping controller
  and observer gain functions for reaction-diffusion {PDE}s,'' \emph{arXiv
  preprint arXiv:2303.10506}, 2023.

\bibitem{lamarque2024adaptive}
M.~Lamarque, L.~Bhan, Y.~Shi, and M.~Krstic, ``Adaptive neural-operator
  backstepping control of a class of nonlinear time-varying {PDE}s,''
  \emph{arXiv preprint arXiv:2401.07521}, 2024.

\bibitem{lamarque2024gain}
M.~Lamarque and M.~Krstic, ``Gain scheduling with a neural operator for a class
  of nonlinear time-varying {PDE}s,'' \emph{arXiv preprint arXiv:2401.02511},
  2024.

\bibitem{jie}
J.~Qi, R.~Vazquez, and M.~Krstic, ``Multi-agent deployment in 3-{D} via {PDE}
  control,'' \emph{IEEE Transactions on Automatic Control}, vol.~60, pp.
  891--906, 2015.

\bibitem{krstic2}
A.~Smyshlyaev, E.~Cerpa, and M.~Krstic, ``Boundary stabilization of a 1-{D}
  wave equation with in-domain antidamping,'' \emph{SIAM J. Control Optim.},
  vol.~48, pp. 4014--4031, 2010.

\bibitem{andrey}
A.~Smyshlyaev and M.~Krstic, ``On control design for {PDE}s with
  space-dependent diffusivity or time-dependent reactivity,''
  \emph{Automatica}, vol.~41, pp. 1601--1608, 2005.

\bibitem{krstic4}
------, \emph{Adaptive Control of Parabolic {PDE}s}.\hskip 1em plus 0.5em minus
  0.4em\relax Princeton University Press, 2010.

\bibitem{vazquez}
R.~Vazquez and M.~Krstic, \emph{Control of Turbulent and Magnetohydrodynamic
  Channel Flow}.\hskip 1em plus 0.5em minus 0.4em\relax Birkhauser, 2008.

\bibitem{convloop}
------, ``Boundary observer for output-feedback stabilization of thermal
  convection loop,'' \emph{IEEE Trans. Control Syst. Technol.}, vol.~18, pp.
  789--797, 2010.

\bibitem{Vazquez2014}
------, ``Marcum {Q}-functions and explicit kernels for stabilization of linear
  hyperbolic systems with constant coefficients,'' \emph{Systems \& Control
  Letters}, vol.~68, pp. 33--42, 2014.

\bibitem{nball}
------, ``Boundary control of reaction-diffusion {PDE}s on balls in spaces of
  arbitrary dimensions,'' \emph{ESAIM:Control Optim. Calc. Var.}, vol.~22,
  no.~4, pp. 1078--1096, 2016.

\bibitem{coupled-parabolic}
------, ``Boundary control of coupled reaction-advection-diffusion systems with
  spatially-varying coefficients,'' \emph{IEEE Transactions on Automatic
  Control}, vol.~62, pp. 2026--2033, 2017.

\bibitem{jing-paper}
R.~Vazquez, M.~Krstic, J.~Zhang, and J.~Qi, ``Kernel well-posedness and
  computation by power series in backstepping output feedback for
  radially-dependent reaction-diffusion {PDE}s on multidimensional balls,''
  \emph{Systems \& Control Letters}, vol. 177, p. 105538, 2023.

\bibitem{vazquez-coron}
R.~Vazquez, E.~Trelat, and J.-M. Coron, ``Control for fast and stable
  laminar-to-high-{R}eynolds-numbers transfer in a 2d {N}avier-{S}tokes channel
  flow,'' \emph{Disc. Cont. Dyn. Syst. Ser. B}, vol.~10, pp. 925--956, 2008.

\bibitem{Power-Series}
R.~Vazquez, G.~Chen, J.~Qiao, and M.~Krstic, ``The power series method to
  compute backstepping kernel gains: Theory and practice,'' in \emph{2023 62nd
  IEEE Conference on Decision and Control (CDC)}, 2023, pp. 8162--8169.

\bibitem{wang2023backstepping}
S.~Wang, M.~Diagne, and M.~Krstic, ``Backstepping neural operators for
  2$\times$2 hyperbolic {PDE}s,'' \emph{arXiv preprint arXiv:2303.10506}, 2023.

\bibitem{mathematica}
I.~Wolfram~Research, \emph{Mathematica}.\hskip 1em plus 0.5em minus 0.4em\relax
  Champaign, IL: Wolfram Research, Inc., 2022.

\end{thebibliography}

\end{document}